\documentclass[11pt,a4paper]{article}
\usepackage{jheppub}
\begin{document}
\newcommand{\mathsym}[1]{{}}
\newcommand{\unicode}[1]{{}}
\newcommand{\bea}{\begin{eqnarray*}}
\newcommand{\eea}{\end{eqnarray*}}
\newcommand{\be}{\begin{equation}}
\newcommand{\ee}{\end{equation}}
\newcommand{\bra}[1]{\langle #1 |}
\newcommand{\ket}[1]{| #1\rangle }

\title{Violation of the Landau-Yang theorem from Infrared Lorentz Symmetry Breaking}
\author[a]{M. Asorey,}
\author[b]{A. P. Balachandran,}, 
\author[c]{M. Arshad Momen}
\author[d]{and B. Qureshi}

\affiliation[a]
{ Centro de Astropart\'{\i}culas y F\'{\i}sica de Altas Energ\'{\i}as, Departamento de F\'{\i}sica Te\'orica,
Universidad de Zaragoza, E-50009 Zaragoza, Spain}
\affiliation[b]{ Physics Department, Syracuse University,
Syracuse, New York 13244-1130, U.S.A.}

\affiliation[c]{Dept. of Physical Science, Independent University, Bangladesh(IUB),  Dhaka 1229, Bangladesh\\
and \\
 Center for Computational and Data Sciences(CCDS), Independent University, Bangladesh(IUB),  Dhaka 1229, Bangladesh}
\affiliation[d]{Department of Mathematical Sciences, IBA, Karachi, Pakistan}
\emailAdd{asorey@unizar.es}
\emailAdd{apbal1938@gmail.com}
\emailAdd{arshad@iub.edu.bd}
\emailAdd{babaraqureshi@gmail.com}

\abstract{Lorentz symmetry forbids decays of massive spin-1 particle like the 
$Z^0$ into two massless photons, a result known as the Landau-Yang theorem. But it is known that  infrared effects can break Lorentz invariance.
Employing the construction of Mund et. al. \cite{MRS} which incorporated this Lorentz 
violation, we propose an interaction leading to the decay $Z^0 \rightarrow 2 \gamma$ and study the dependence of the decay on the parameter of this Lorentz violation.}

\arxivnumber{2304.11008}

%\begin{document}
\maketitle

\baselineskip=18pt

\section{Introduction}
In Quantum Field Theories (QFTs)  with a mass gap, it is assumed that the Poincar\'e
group has a unitary action in the Hilbert space of the theory leaving the vacuum invariant. 
It is thus a `symmetry' of the theory acting on the local observables in a bounded 
spacetime region. Causality can also be formulated by requiring the local observables
in spacelike complements to commute.

But there are physically important theories with no mass gap, QED being a 
prime example. In QED, because of infrared effects, there are uncountably many 
superselection sectors, and Lorentz transformations map one such sector into another. 
 Hence Lorentz symmetry is spontaneously broken \cite{strocchi}, just like the U(1) gauge group which is spontaneously broken in superconductivity or ferromagnetism.
 
 There is an important theorem of Landau and Yang  \cite{Landau}, \cite{Yang} for Lorentz invariant theories : 
 the decay of a massive spin 1 particle such as $Z^0$ to two photons is forbidden. This
 theorem is remarkable: its proof does not rely on principles of quantum field theory (QFT) such as causality.
This result is due to the fact that the the Clebsch-Gordan series for the tensor
product of two  irreducible Poincar\'e group representations of photons,
when Bose symmetrised, does not contain a massive spin one representation, as shown {\color{blue} by} Balachandran, Jo and Marmo \cite{BalJoMarmo}. There are 
generalisations of this result to other decays \cite{joBal} shown there. For example, $Z^0$ cannot decay into two massless neutrinos. 

Thus, the Landau-Yang and related theorems use only Poincar\'e
invariance and statistics of identical particles and invokes no other principles of quantum field theory. There are also experiments in non-linear optics which put limits on the rate of such Landau-Yang selection rule violating processes \cite{Koslov, English, Plunien}. The observation of any one of these decays will thus profoundly affect QFT.

On the other hand, it has been known for a long time that infrared effects in QED create
uncountably many superselection sectors and that the action of Lorentz transformations
interchanges these sectors. Hence by definition, Lorentz symmetry is spontaneously broken.
In a recent paper, Mund, Rehren and Schroer \cite{MRS} have reviewed these results and 
 developed a theory of ‘infrafields’ which can serve as order parameters for this symmetry breaking. Given these results and developments, it is natural to ask if a model for say $Z^0 \rightarrow 2 \gamma$ decay can be formulated using the infrafields. In this paper, we argue that this can be done and an explicit decay rate can be derived. We emphasize that the mechanism of $Z^0 \rightarrow 2\gamma$  decay that we are discussing does not invoke any Beyond Standard Model (BSM) physics as {\color{blue} happens} with extra generations of fermions ( see e.g. \cite{Peskin}) or by requiring explicit Lorentz violating new interactions ( see e.g. \cite{Kost} for a recent review ). Rather it arises because of the infrared structure of QED itself.

In the subsequent sections, we review the construction of infrafields and their response to gauge transformations.
Then we formulate an interaction for $Z^0 \rightarrow  2 \gamma$  which is invariant under ``small" or Gauss law-generated gauge transformations which is not Lorentz invariant and compute the rate for the above decay.

\section{The Infrafield}

Such fields first acquired a prominent role in the quantum field theory of massless
`continuous spin' particles which occur in the work of Wigner\cite{weinberg}. Their momenta
are lightlike and future pointing so that no known physical principle forbids their existence. 
But Yngvason \cite{Yngvason} first proved that local quantum fields do not exist for such particles.
It was later shown that ‘fields localised on cones’ do exist for these particles.  These fields were later adapted to QED
for its formulation entirely on a Hilbert space, avoiding indefinite metric 
constructions altogether. They relied on the axial gauge and Dirac’s construction of 
gauge invariant fields. 

The Wilson line from $x$ to $\infty$ in the spacelike direction $\eta$
gives the infrafield $\phi$, the above Wilson line for charge $q$ being $e^{iq\phi(x,\eta)}$.
Then,  under a $U(1)$ gauge transformation $e^{i\chi}$, the Wilson line 
\[\
W(x,\eta) = e^{iq \int_0^\infty  A_\mu (x+ \tau \eta) \eta^\mu d\tau} \equiv e^{iq \phi(x,\eta)} 
\]
transforms as
\[
W(x,\eta) \rightarrow  e^{i q\chi(\eta_\infty )} W(x,\eta) e^{-iq \chi(x)}, \qquad  \eta_\infty  := \lim_{\tau\rightarrow \infty} \tau \eta.
\]
This transformation suggests the interpretation that $W$ carries a charge, say $-q$ at $x$ and $q$ at $\eta_\infty $.
If $\psi$ is a charge $q$ field, then following Dirac\cite{Dirac} and Mandelstam\cite{Mandelstam}, we see that $W(x, \eta) \psi(x)$ is small gauge invariant, but still has a charge $q$ blip at $\eta_\infty $.

If $\psi$ is a field of charge $q$ , then as Dirac observed, the field $e^{iq\phi(x,\eta)}
\psi (x)$ is invariant under local or small gauge transformations.
It can be smeared in $x$ with test functions localised in a finite region containing $x$ without spoiling small gauge invariance. But acting on the vacuum, it produces only  a charge $q$ blip at infinity. So it produces surface excitations at infinity. That is the case even if $x$ is smeared.

We want to choose $\eta$ so that $A_\mu \eta^\mu$ does not produce negative norm states acting on the vacuum.
That requires that this field has only spacelike 
components in $\mu$. Hence $\eta^\mu$ is chosen to be along a  spacelike direction. In the mostly minus metric convention, it follows that $\eta.\eta = - \kappa^2$, with $\kappa$ real. 
But unlike in the axial gauge, we do not fix the  direction  of $\eta$.

For fixed $\kappa$, this is a deSitter space with a spherical boundary for finite $\eta$.
But  $W (x, \eta) $ is invariant under the scalings $\eta \to  c \eta, c>0$. So we can fix $\kappa$ so that a compactification with a spherical boundary is a preferred choice of spacelike
boundaries.

Under standard Lorentz transformations $\Lambda$ of $A_\mu$ , not only the argument, 
but also the index $\mu$ of $A_\mu$ gets transformed, so that 
\[
\Lambda :\phi (x, \eta) \rightarrow \phi (\Lambda x, \Lambda \eta).
\]
Hence, $\eta$ as well gets transformed in the ``escort" field $\phi(x,\eta)$. 

\section{ A Model Interaction for $Z^0 \rightarrow 2 \gamma$} 
We need the interaction to have the following properties :
\begin{enumerate}
\item It should be gauge invariant under small gauge transformations. 
\item The net charge at infinity should add up to zero so that charge conservation
is maintained. So we want to preserve global $U(1)$ invariance. 
\end{enumerate}

For the escort field of charge $q$, the operator 
\[ 
: e^{i q\phi(x,\eta)}:\quad : e^{ -iq\phi(x,\eta’) }:
\]
is invariant under small gauge transformations and has zero net charge. 
Also  $Z^0 _\mu (x)$ is neutral under this $U(1)$. Hence if $f^\mu$ is a vector-valued test function, 
a Gauss law-invariant interaction with zero net charge is
\begin{equation}
 d ~f^\mu (x)  Z^0_\mu (x): e^{i q\phi(x,\eta)}:\quad : e^{ -iq\phi(x,\eta’) }:
 \label{interaction}
\end{equation}
with an interaction strength $d$. 

\begin{figure}[hbtp]
\centering
\includegraphics[width=.7\textwidth]{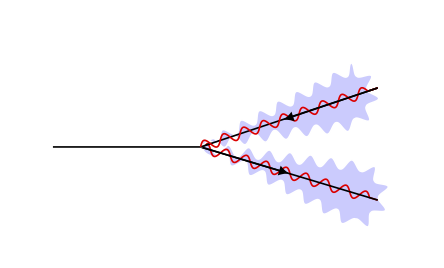} 
 \caption{Diagram associated to the decay of $Z^0$ into two  $\gamma$s in the presence of escort fields.}
\label{ffig1}
\end{figure}

A natural choice for $f^\mu (x)$ is $h^\mu $ where $h^\mu$ is a 
chosen polarisation vector for $Z^\mu$ in its rest system. This choice is suggested 
as we will calculate the decay in the $Z^\mu$ rest system.
We can also smear just $Z^0_\mu$ with a scalar function $ f(x)$ and use a fixed vector
$h^\mu$ for this polarization vector. But our preferred choice in this paper is (\ref{interaction}).
Finally we have also the option of smearing the variables  $\eta$ and $\eta’$ , but we will avoid it for now. 
The process which lets us avoid the Landau-Yang theorem by means of the escort fields is sumarized in
 Figure \ref{ffig1}.

\section{Calculation of the Amplitude using the Mode Expansion of the Escort Field}
We can obtain the mode expansion of $e^{i \phi}$ from that of $A_\mu$ ( following Mund et. al. \cite{MRS})  :
\[
A_\mu(x) = \int \tilde{d}k \left[a_\mu(k) e^{-ik \cdot x} + a^\dagger_\mu(k) e^{ik \cdot x} \right] \]
with the Lorentz invariant phase space (LIPS) measure: 
\[ \tilde{d}k= \frac{d^3k}{(2\pi)^3 ~2 |\mathbf{k}|}.
\]

Although the operator $A_\mu(x)$ is defined only on an indefinite metric (Krein) space , 
that is not the case for $A_\mu \eta^\mu$ as $\eta^\mu$ is spacelike.

One then defines 
\[
\phi(x,\eta) = \int_0^\infty d\tau A_\mu(x + \eta \tau)\eta^\mu d\tau.  
\]

Expanding $\phi(x)$  ( using (+,-,-,-) metric), one gets, using the  ``$i \epsilon$" prescription,
\bea
\phi(x,\eta) &=& \int \tilde{d}k \eta^\mu \left[ a_\mu(k) e^{-i k\cdot x}  \int_0^\infty d\tau~ e^{-i [(k. \eta) - i\epsilon] \tau  } \right.  \\
&+& \left. a^\dagger_\mu(k) e^{i k\cdot x}  \int_0^\infty d\tau~ e^{i [(k. \eta) + i\epsilon] \tau  } 
\right] \\
&=& i \int \tilde{d}k ~\eta^\mu \left[ \frac{a_\mu(k)}{\eta \cdot k - i\epsilon}  e^{-i k\cdot x}   - \frac{a^\dagger_\mu(k)}{\eta \cdot k + i\epsilon}  e^{i k\cdot x}  
\right] \\
&=& i \int \tilde{d}k  \left[ \chi_k(\eta)  e^{-i k\cdot x}   - \chi^\dagger_k(\eta)  e^{i k\cdot x} \right]
\eea
where $$\chi_k(\eta)  \equiv \frac{a(k)\cdot \eta}{k.e - i \epsilon}$$. 

Therefore 
\bea
\phi(x,\eta) &=& i  \left[\int  \tilde{d}k  \left( \chi_k(\eta)  e^{-i k\cdot x}   - \chi^\dagger_k(\eta)  e^{i k\cdot x} \right) \right] \\
%& \equiv&
%i \left[\int  \tilde{d}k  \left( \chi_{F,k}  e^{-i k\cdot x}   - \chi^\dagger_{F,k}  e^{i k\cdot x} \right) \right]
\eea
%where 
%\[
%\chi_{F,k} \equiv \int De F(e) \chi_k(e).
%\]

Now 
\[
:e^{i q \phi(x,\eta)}: =  
e^{q \int \tilde{d}k  \chi^\dagger_{k} e^{ik \cdot x} }  e^{-q \int \tilde{d}k \chi_{k} e^{-ik \cdot x}  }
% 
%e^{\frac{q^2}{2} \int \tilde{d}k \int d\tilde{k'} [ \chi_{F,k}, \chi^\dagger_{F,k'} ] e^{-i(k-k').x}}
\]

Accordingly
\bea
 :e^{iq \phi(x,\eta)}&:~:&e^{-iq \phi(x,\eta')}: =  \\
&~&e^{q \int \tilde{d}k  \chi^\dagger_{k} e^{ik \cdot x} }  e^{-q \int \tilde{d}k \chi_{k} e^{-ik \cdot x}  }e^{-q \int d\tilde{k'}  \chi^\dagger_{k'} e^{ik' \cdot x} }  e^{q \int d\tilde{k'} \chi_{k'} e^{-ik' \cdot x}  }
\eea
As Baker-Campbell-Haussdorf formula gives us  
\[
e^M e^N = e^{M+N +\frac{1}{2} [ M,N]+\cdots} \Rightarrow e^M e^N = e^N e^M e^{[ M,N]}
\]
where in the last step we have assumed $[M,N]$ is a $c$-number,

\begin{eqnarray}
 &&:e^{iq \phi(x,\eta)}::e^{-iq \phi(x,\eta')}: =  \nonumber \\
&&e^{q \int \tilde{d}k  \chi^\dagger_{k} e^{ik \cdot x} } e^{-q \int d\tilde{k'}  \chi^\dagger_{k'} e^{ik' \cdot x} } e^{-q \int \tilde{d}k \chi_{k} e^{-ik \cdot x}  }  e^{q \int d\tilde{k'} \chi_{k'} e^{-ik' \cdot x}  } e^{\Delta(\eta,\eta',x)}.
\label{cnumber}
\end{eqnarray}
Now 
\be
\Delta(\eta,\eta',x) =q^2 \int \int \tilde{d}kd\tilde{k'} [\chi_{k} , \chi^\dagger_{k'} ]e^{-i(k-k')\cdot x} 
\ee
But
\be
[\chi_{k}, \chi^\dagger_{k'} ] =  [ \chi_k(\eta), \chi^\dagger_{k'}(\eta')] = ( e \cdot e')\frac{(2\pi)^3 2E_k \delta^3(k-k') }{(k.\eta + i\epsilon)(k'.\eta' - i\epsilon)}\\
\ee

 Hence,
 \bea
 \Delta(\eta,\eta',x) =&q^2&  ( \eta\cdot \eta')\times \\
  &~&\left\{ \int \int \tilde{d}k \tilde{d}k' (2\pi)^3 2E_k \delta^3(\vec{k}-\vec{k}')  \frac{e^{-i(k-k')\cdot x} }{(k.\eta + i\epsilon)(k'.\eta' - i\epsilon)} \right\} \\
   \!\! &~&=q^2 \int \tilde{d}k \frac{1}{(k.\eta + i\epsilon)(k.\eta' - i\epsilon)} \equiv q^2 Q(\eta, \eta')
\eea 
ends up being independent of $x$. Note that 
in the last step,  we have carried out the $k'$ integration. 

%Let us next focus on the integral: 
%\[
%Q(\eta, \eta') \equiv \int \tilde{d}k \frac{1}{(k.\eta + i\epsilon)(k.\eta' - i\epsilon)}
%\]
%( Note: The $\tilde{d}k$ integration is over the three dimensional $k$ and also $e$ is a spacelike 
%vector).
 
%As $k`.k =0  \mbox{~or}~ \mu^2$, 
The integral $Q(\eta, \eta')$ can be rewritten ( using Feynman-Schwinger parameterization ) as
\bea
Q(\eta, \eta')= \int \tilde{d}k\int_0^1 d\lambda \frac{ ( \eta\cdot \eta')}{[ \lambda ( k\cdot \eta + i\epsilon ) + ( 1-\lambda)  ( k\cdot \eta' - i\epsilon ) ]^2} \\
=  \int \tilde{d}k\int_0^1 d\lambda \frac{ ( \eta\cdot \eta')}{[  k\cdot \eta' + \lambda  k.( \eta-\eta') - i\epsilon ( 1-2\lambda)  )  ]^2} \\
= \frac{ ( \eta\cdot \eta') }{2(2\pi)^3} \int_0^\infty k dk~\int_0^1 d\lambda \int d\Omega_k \frac{1}{\left[k.( \eta' + \lambda( \eta-\eta'))+ i \epsilon (1-2\lambda)\right]^2} .
\eea

%Using 
%\bea 
%I(e, e') = (e. e')  \int \tilde{d}k\int_0^1 dx \frac{1}{[ x ( k\cdot e - i\eta ) + ( 1-x)  ( k\cdot e' + i\eta ) ]^2} \\
%= \frac{1}{2}  (e. e')  \int \tilde{d}k\int_0^1 dx \frac{1}{[  k\cdot e' + x  k.( e-e') + i\eta ( 1-2x)  )  ]^2} \\
%= \frac{1}{2}  (e. e') \int_0^\infty k dk~\int_0^1 dx \int d\Omega_k \frac{1}{\left[k.( e' + x( e-e'))+ i \eta (1-2x)\right]^2}
%\eea

Using the identity 
\[
\int d\Omega_{\hat{n}} \frac{1}{(\hat{n} \cdot \vec{a} + b)^2} = \frac{4\pi}{( b^2 - \vec{a}\cdot\vec{a})},
\] 
we  get 
\bea
&&Q(\eta,\eta') =%  \\
%&&\frac{1}{(2\pi)^2} (\eta. \eta') \int_0^\infty \frac{1}{k} dk~\int_0^1 \frac{d\lambda}{[(1-2\lambda)^2\epsilon^2 + ( 1+2\lambda^2 ( 1- (\eta.\eta')) + 2\lambda ( \eta.\eta' -1))^2}\\ 
 %&=& 
  \frac{(\eta. \eta')}{ (2\pi)^2}  \int_0^\infty \frac{1}{k} dk~\int_0^1 \frac{d\lambda}{[(2\lambda^2 -2\lambda+1) + 2\lambda(1-\lambda)\eta.\eta' ]^2 }
\eea
where the limit $\epsilon \rightarrow 0_+$ has been employed. The $k$ integral is logarithmically divergent and  needs both an 
 ultraviolet and infrared momentum cutoff,  $M_{UV}$ and $m_{IR}$ respectively: 
\bea
 Q(\eta,\eta') &=&  \frac{(\eta. \eta')}{ (2\pi)^2} \ln\left(\frac{M_{UV}}{m_{IR}}\right)\int_0^1 \frac{d\lambda}{\left[(2\lambda^2 -2\lambda+1) + 2\lambda(1-\lambda)\eta.\eta' \right]^2 } \\
&\equiv&\frac{\ln\left(\frac{M_{UV}}{m_{IR}} \right)}{ (2\pi)^2} S( \eta\cdot \eta')
\eea
%{\color{red}
Note that the UV part of the logarithmic divergences of $\int \frac{1}{k} dk$  can alternatively be absorbed
by the renormalization of the electric charge: 

\[
  \Delta = \frac{3% (\eta \cdot \eta')
 }{2}\left(1-\frac{q^2}{q_R^2}\right) S(\eta \cdot \eta')=\frac{3 %(\eta \cdot \eta')
 }{2}\left(1-\frac{4 \pi}{137 q_R^2}\right) S(\eta \cdot \eta')
\]
%}
%\hrule\hrule 
%\vspace{16pt}
%
%\label{integral}
Thus the $c$-number factor appearing in (\ref{cnumber}) takes the form
\be
e^{\Delta(\eta, \eta')} = e^{\frac{3 
 }{2}\left(1-\frac{4 \pi}{137 q_R^2}\right) S(\eta \cdot \eta')} %(\eta \cdot \eta')}
%\left(\frac{M_{UV}}{m_{IR}}\right)^\frac{q^2 S(\eta\cdot \eta')}{2\pi^2}
\label{factor}
\ee
where we have dropped the $x$ argument in $\Delta(\eta, \eta',x)$ as it is independent of $x$. 

%%%%%%

\begin{figure}[hbtp]
\centering
\includegraphics[width=.9\textwidth]{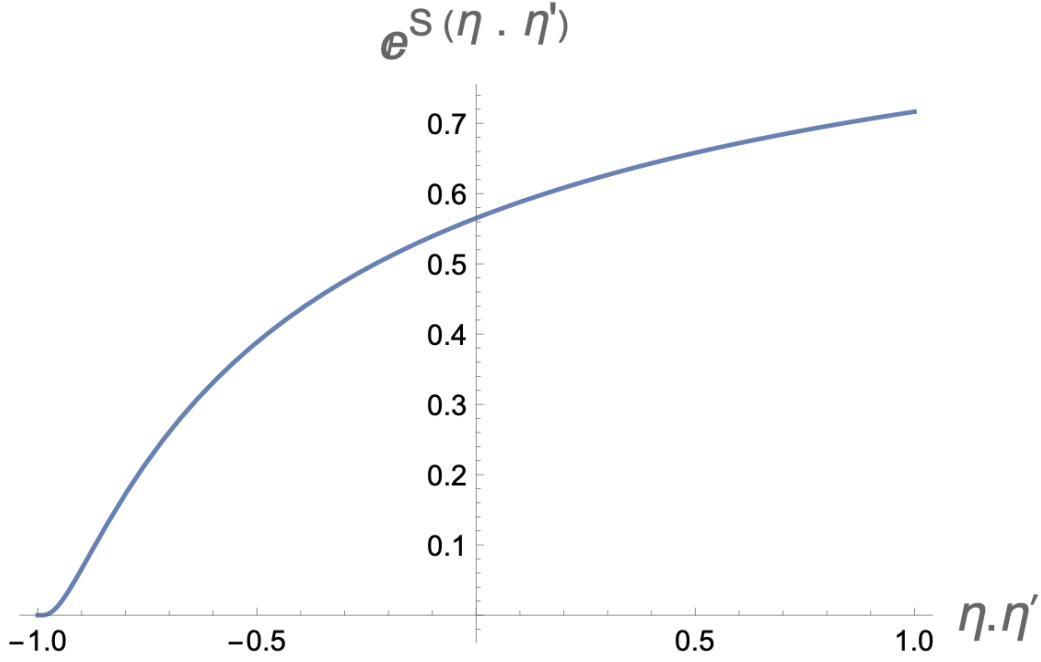}
\caption{The plot of the $e^{S(\eta, \eta')}$ vs $(\eta\cdot \eta')$}
\label{fig2}
\end{figure}

A plot of the function $e^{S(\eta, \eta')} $ is shown in Figure \ref{fig2}
Note that when $\eta = \eta'$ so that $\eta \cdot \eta'=-1$ , the factor $e^\Delta$
%$$e^\Delta=  \left(\frac{M_{UV}}{m_{IR}}\right)^{-\frac{q^2}{12\pi^2}}$$
 vanishes. That is,  as long as we remain in the same ``superselection" sector,  the Landau-Yang theorem is validated. But as long as $\eta \neq \eta'$ there is a none-zero probability for the decay of $Z^0$ into two photons. 

Let us recall that the decay width of a  particle of mass $m_A$ at rest into two identical massless particles is given by the formula:
\be 
\Gamma = \frac{1}{32\pi m_A^2} |\mathcal{A}|^2.
\label{gamma}
\ee  
%
% \begin{figure}[hbtp]
%\centering
%%\includegraphics[scale=0.5]{angular_distribution.pdf}
%\end{figure}

Now we are interested in the product of matrix elements : 
 \bea
&~&\bra{k_1, \varepsilon_1; k_2, \varepsilon_2}  :e^{iq \phi(x,\eta)}: : e^{-iq \phi(x,\eta')}: \ket{0}_\gamma \bra{0} f\cdot Z\ket{Z, h}_Z \\
 &=& -q^2  e^{\Delta(\eta, \eta')} \int  \int  \tilde{d}k  d\tilde{k'} \bra{k_1, \varepsilon_1; k_2, \varepsilon_2} \chi^\dagger_{k} \chi^\dagger_{k'} \ket{0}_\gamma  \times  \bra{0} f \cdot Z \ket{Z, h} \int d^4 x  e^{ik \cdot x}   e^{ik' \cdot x} e^{-i P.x} \\
 &=& -q^2  e^{\Delta(\eta, \eta')} \left[ \frac{( \eta \cdot \varepsilon_1)}{(\eta \cdot k_1 + i \epsilon)}  \frac{( \eta' \cdot \varepsilon_2)}{(\eta' \cdot k_2 + i \epsilon)} + 
 \frac{( \eta \cdot \varepsilon_2)}{(\eta \cdot k_2 + i \epsilon)}  \frac{( \eta' \cdot \varepsilon_1)}{(\eta' \cdot k_1 + i \epsilon)}    \right] 
 (f \cdot h) 
\eea
where $\varepsilon_{1,2}$ are the polarization vectors of the photons with the momenta $k_1, k_2$ respectively, while $h$ is the polarization/spin vector associated with the $Z_0$ boson.

%Now
% \bea
% e^{iq \phi(x,e)} e^{-iq \phi(x,e')} &=& e^{q \int \tilde{d}k ` \chi^\dagger_{F.k} e^{-ik \cdot x} }  e^{-q \int \tilde{d}k \chi_{F,k} e^{-ik \cdot x}  } e^{-q \int \tilde{d}k \chi^\dagger_{F'.k} e^{-ik \cdot x} }  e^{q \int \tilde{d}k \chi_{F',k} e^{-ik \cdot x}  } \\
% &=&  e^{q \int \tilde{d}k ` \chi^\dagger_{F.k} e^{-ik \cdot x} }  e^{-q \int \tilde{d}k \chi^\dagger_{F',k} e^{ik \cdot x}  } e^{-q \int \tilde{d}k \chi_{F'.k} e^{-ik \cdot x} }  e^{q \int \tilde{d}k \chi_{F',k} e^{-ik \cdot x}  } e^{-q^2 /2 \int \tilde{d}k \int d\tilde{k'} [ \chi_{F, k'} , \chi^\dagger_{F,k'} e^{i(k-k')\cdot x}} 
% \eea

%{\color{red}
%Note that the UV part of the logarithmic divergences of $\int_0^\Lambda \frac{1}{k} dk$  can alternatively be absorbed
%by the renormalization of the electric charge: 
%
%\[
% q^2 \Delta = \frac{3 \eta \cdot \eta'}{2}\left(1-\frac{q^2}{q_R^2}\right)=\frac{3 \eta \cdot \eta'}{2}\left(1-\frac{4 \pi}{137 q_R^2}\right)
%\]
%
%%%%%$\frac{q_{ren}^2}{4 \pi}\approx \frac1{\frac{1}{6 \pi} q^2\log{\Lambda/\mu}}$
%%$$q^2 \log{\Lambda/\mu}\approx \frac{24 \pi^2}{q_{ren}^2}$$
%}
    %B) Completion of the calculation of $\Gamma$ as there is problem with the use of the formula (\ref{gamma}) which breaks the isotropy due to the presence of three vectors $\eta, \eta', f$ even after initial averaging and the summing of the polarization in the final states. Is a  full calculation from first principle needed?   \\

The above result confirms the claim that escort field can allow decays which are forbidden in the Standard Model. In particular the decay of {\color{blue}$Z^0$} into two photons is now possible for any value of $\eta\cdot\eta'$ except for $\eta\cdot\eta'=-1$ where we recover  the standard Landau-Yang result. The reason being that in this case the two branches of escort field  cancel out.

Decay of $Z^0$ into two photons could in principle be observable in collider experiments. Observation of such an event will have profound consequences and can be interpreted as evidence of the novel infrared structure of QED. On the other hand, limits on decay width of this process from collider data can constrain the Lorentz violation that we have discussed in the paper. 
  
Violation of the Landau-Yang theorem by the inclusion of the 
escort field allows for study of  other interesting physics phenomena. Not only one has the possibility of studying novel and rare decays in atomic physics but also its impact in the cosmology for instance in the mapping of the 21cm hydrogen line. We hope to address to some of this phenomena using this framework in the future.

\section*{Appendix}

We give a closed expression for the integral encountered in the text :
\[
F({\eta\cdot \eta'}){:=}\int_0^1 \frac{1}{\left(2 ({\eta\cdot \eta'}) (1-\lambda) \lambda+\left(2 \lambda^2-2 \lambda+1\right)\right)^2} \, d\lambda
\]

This integration can be done using %Mathematica which gives: 
the identity
\[
\int_0^1 \frac{d\lambda}{[(2\lambda^2 -2\lambda+1) + 2\lambda(1-\lambda)\eta\cdot\eta' ]^2 } = \frac{1}{1+\eta\cdot\eta'} + 
\frac{2 \tan^{-1} \left( \sqrt{\frac{1-\eta\cdot\eta'}{1+\eta\cdot\eta'}}\right) }{(1+\eta\cdot\eta') \sqrt{1-(\eta\cdot \eta')^2}}
\]

Parameterizing  $A= \eta \cdot \eta'= - \cos \Theta $ (as both of them are spacelike vectors), one gets a simplified expression 
\[
F(\Theta) = \frac{ 1 + \frac{\Theta}{\sin \Theta}}{1-\cos \Theta}
\]``
%\text{Plot}\left[(A F(A))^2,\{A,-1,1\}\right]
%
%\text{Null}
\section*{Acknowledgements}

M.A. is  partially supported by Spanish MINECO/FEDER Grant PGC2022-126078NB-C21 funded by MCIN/AEI/10.13039/501100011033,  DGA-FSE grant 2020-E21-17R and Plan de Recuperación, Transformación y Resiliencia - supported European Union – NextGenerationEU. We wish express our thanks for Amilcar Queiroz for discussions on the early stages of this investigation. 

\newcounter{mathematicapage}

%\subsection*{Mathematica Code} 
%\begin{doublespace}
%\noindent\(\pmb{F[\text{A$\_$}]\text{:=} \text{Integrate}[1/((2x{}^{\wedge}2 -2x +1) + 2x(1-x)*A){}^{\wedge}2,\{x,0,1\}]}\\
%\pmb{\text{Plot}[(\text{Exp}[A F[A]]){}^{\wedge}2,\{A,-1,1\}]}\\
%\pmb{}\)
%\end{doublespace}
%
%
%\begin{figure}[hbtp]
%\caption{The square of the Amplitude (\ref{integral})}
%\centering
%\includegraphics[scale=1]{Plot_2.pdf}
%\end{figure}
%Question: Which one to incorporate? 
%\newpage
--------------------------------------
%\section*{Things to do now:}

%1. Arshad would write some closing comments. Not done yet.\\

\end{document}